\documentclass[aps, reprint, groupedaddress, longbibliography]{revtex4-2}
\usepackage{graphicx, amsmath, mhchem}
\usepackage{subfigure, amssymb}
\usepackage{float} 
\usepackage[makeroom]{cancel}
\usepackage{color}

\begin{document}

\title{Critical behavior in the Artificial Axon}
\author{Ziqi Pi}
\author{Giovanni Zocchi}
\email{zocchi@physics.ucla.edu}
\affiliation{Department of Physics and Astronomy, University of California - Los Angeles}

\begin{abstract}

\noindent The Artificial Axon is a unique synthetic system, based on biomolecular components, which supports 
action potentials. Here we examine, experimentally and theoretically, the properties of the threshold for firing 
in this system. As in real neurons, this threshold corresponds to the critical point of a saddle-node bifurcation. 
We measure the delay time for firing as a function of the distance to threshold, recovering the expected scaling 
exponent of $- 1/2$. We introduce a minimal model of the Morris-Lecar type, validate it on the experiments, and 
use it to extend analytical results obtained in the limit of "fast" ion channel dynamics. In particular, we discuss the dependence of the firing threshold on the number of channels. 
The Artificial Axon is a simplified system, an Ur-neuron, relying on only one ion channel species for functioning. 
Nonetheless, universal properties such as the action potential behavior near threshold are the same as in real neurons. 
Thus we may think of the Artificial Axon as a cell-free breadboard for electrophysiology research.

\end{abstract}

\maketitle

\section {Introduction}

\begin{figure}
\includegraphics[width=\linewidth]{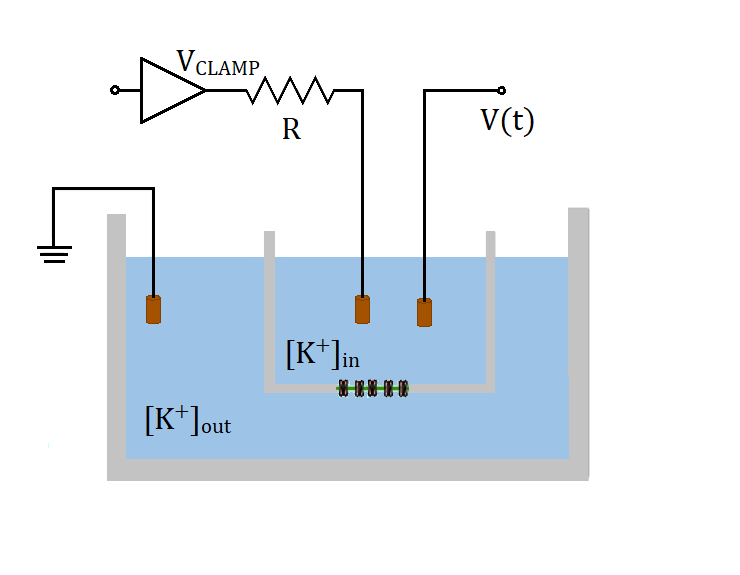}
\caption{Schematic drawing of the Artificial Axon. The $\sim 1 \, mL$ volume "inside" compartment has 
a lower $KCl$ concentration ($[KCl]_{in} \approx 30 \, mM$) compared to the outside compartment 
($[KCl]_{out} \approx 150 \, mM$). A $100 \, \mu m$ size phospholipid membrane patch ({\it not} drawn to scale) 
with embedded ion channels separates the two compartments. We refer voltages to the (grounded) outside 
chamber. Voltages are read and currents are injected through $AgCl$ electrodes. Also shown is the 
Current Limited Voltage Clamp (CLVC), consisting of a voltage clamp plus series resistance $R$. 
The electrode on the right is used to measure the axon voltage $V(t)$, through a high impedance pre-amp. 
The electronics is controlled by a Labview program. }
\label{fig:setup}
\end{figure}

\noindent Action potentials are an interesting dynamical system, but not easy to study
due to the complexity of the neuron. We recently introduced the idea of producing 
action potentials in vitro; our cell free system is based on the reconstituted biological 
components (phospholipid membrane, voltage gated ion channels) and thus on the same microscopic 
mechanism for generating voltage spikes as real neurons; we call it the Artificial Axon (AA) 
\cite{Amila2016, Hector2017}.  While our ultimate goal is to study the behavior  of simple networks 
of Artificial Axons (an initial demonstration is given in \cite{Hector2019}), it is also interesting to develop the AA 
as a cell free platform for electrophysiology studies. Specifically, our hope is that AA platforms may one day 
replace, in the laboratory, cultured neurons, which typically come from rats. In addition, this experimental system 
provides motivation and inspiration for developing models, numerical experiments, 
and eventually theory in the general field of complex systems. 
In this context, we concentrate here on one feature of an individual AA, which has the allure of universality 
(and thus is relevant for real neurons too), but which is difficult to measure in real neurons, namely 
the behavior near threshold for firing an action potential. In the language of dynamical systems, the threshold 
for firing is related to the critical point of a saddle node bifurcation. This correspondence has of course 
been explored before in neurons 
\cite{Morris_Lecar, Izhikevich2000, Tsumoto2006, Sejnowski2008, Zhao2017, Koch_Book}. 
Notwithstanding, the AA is a simpler system 
compared to a real neuron, both in practice and conceptually. It is a dressed down experimental system 
where we try to distill the minimal components which can still generate an action potential. Notably, it has 
only one ion channel species and relies on one ionic gradient across the membrane which is the active element 
of the system. The role of a second ionic species and channel type is played instead by a simple electronic 
device which we call a current limited voltage clamp (CLVC) \cite{Amila2016}. The command voltage to this clamp 
is a control parameter for the dynamical system slightly different from the control parameters in traditional 
electrophysiology. As a function of this control parameter the system exhibits critical slowing down and scaling 
of the time to fire. \\
Our purpose with this paper is to use the AA as a conceptual tool - a model - to generate the simplest interesting 
dynamical systems related to action potentials. We start from some experimental measurements of critical 
behavior in the physical AA and then move on to discuss this critical behavior for some corresponding dynamical 
systems. But first, let us introduce the AA, which we have described before \cite{Amila2016, Hector2017}. 
It is a black lipid bilayer about $200 \, \mu m$ in size, separating two compartments ("in" and "out") 
with ionic solutions at different salt concentrations (and matched osmotic pressure), 
see Fig. \ref{fig:setup}. 

\begin{figure}
\includegraphics[width=\linewidth]{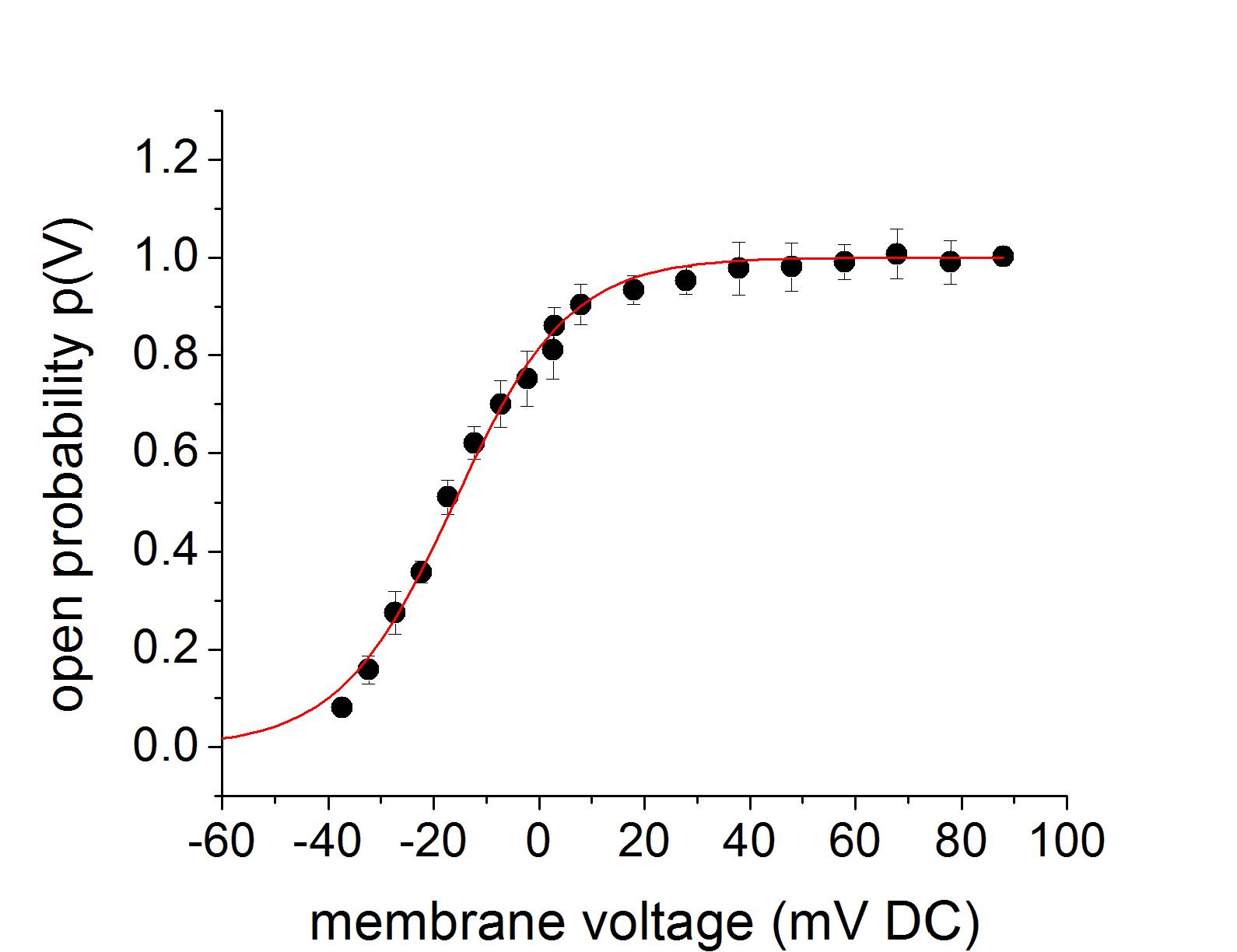}
\caption{Equilibrium open probability measured for the KvAP channels in the AA setting. 
Channels are closed below $\sim - 50 \, mV$ and open at positive voltages. This display 
does not take into account inactivation (see text). 
Reproduced from \cite{Amila2013}.}
\label{p_open}
\end{figure}

Typically we keep the inside chamber at a salt concentration $[KCL] = 30 \, mM$ and the outside at 
$[KCL] = 150 \, mM$. Given a small leak conductance for $K^+$ ions across the membrane (which is 
impermeable to other ionic species), equilibrium with respect to ionic currents is reached for a positive Nernst 
potential inside, with respect to the grounded outside: 

\begin{equation}
V_N = \frac{k T}{|e|} ln \frac{[K^+]_{out}}{[K^+]_{in}} \, \approx \, + \, 50 \, mV
\label{eq: Nernst}
\end{equation}

\noindent where $|e|$ is the charge of the $K^+$ ion, $T$ the temperature, $k$ the Boltzmann constant, and square 
brackets denote concentration. Embedded in the phospholipid membrane are order of 100 voltage gated 
potassium ion channels: these are the active elements of the system. Voltage gated ion channels are the 
ionics equivalent of transistors in electronics; they are membrane proteins which act as pores for one specific 
ion species (potassium in our case), are impermeable to other species, and switch from closed 
(not conducting) to open (conducting) depending on the voltage across the membrane. We use an Archaean channel 
(KvAP) which can be expressed in E. coli and reconstituted in vesicles \cite{schmidt_gating_2009}, which are then fused 
to the AA membrane (see Mat. $\&$ Met.). The channel is closed for voltages below about $- 50 \, mV$ and opens 
at positive voltages; its main characteristics are that it is slow (opening rates at positive voltage are of order 
$\sim 25 \, ms$ \cite{Amila2016}) and that at positive voltage it eventually enters an "inactive" state which is functionally 
closed, but from which the open state is not accessible. Recovery from this inactive state necessitates 
negative voltages ($< - 100 \, mV$) and is also slow (measured in the hundreds of ms). The detailed channel 
dynamics are complex \cite{schmidt_gating_2009}, but we will not need to enter into all the details for this study. 
Fig. \ref{p_open} shows the measured open probability curve $p_o(V)$ for the channels in our system, 
reproduced from \cite{Amila2013}. The meaning of this plot is that it represents the equilibrium probability that 
the channel is open if the voltage was held steady at $V$, and if the inactive state was not there. 
The plot is obtained by quickly stepping the voltage to $V$ (using a voltage clamp) from a large negative value 
where all the channels are in the closed state, and measuring the peak current (the current through the channels 
before they inactivate). \\ 
The last component of the AA is the current limited voltage clamp (CLVC). This simple device consists of 
a traditional voltage clamp with a large resistance in series (Fig. \ref{fig:setup}). Given a voltage $V(t)$ 
in the AA (in the inside chamber, relative to the grounded outside chamber), a command voltage 
$V_{clamp}$ to the CLVC causes a current $I_{clamp} = [V_{clamp} - V(t)] / R$ to be injected in the AA. 
The role of the CLVC is to hold the system off equilibrium at a negative "resting potential" $V_r$, while still 
allowing the voltage in the AA to fluctuate in response to perturbations. We can now write down the equation 
governing the voltage in the AA (basically the same equation is the starting point for all models of the 
neuron \cite{Koch_Book}): 

\begin{equation}
\frac{d V}{d t} = \frac{N_0 \chi}{C} \, [ p_o(t) + \chi_{\ell} / \chi ] [V_N - V(t)] +  
\frac{1}{RC} \, [V_{clamp} - V(t)]
\label{eq: membrane}
\end{equation}

\noindent $C$ is the membrane capacitance, $N_0$ the number of ion channels, $\chi$ the single channel 
conductance (with the channel open), $\chi_{\ell}$ a small leak conductance which is present even if the 
channel is closed ($\chi_{\ell} << \chi$), $p_o(t)$ the probability that the channel is open, $V_N$ the Nernst 
potential (\ref{eq: Nernst}), $R$ the series resistance of the CLVC, and $V_{clamp}$ the command voltage 
to the CLVC. If we multiply both sides by $C$, the equation describes the charging of the membrane capacitance 
by the currents specified on the RHS: the current through the channels, with driving force proportional to 
$[V_N - V(t)]$, and the current injected by the CLVC, with driving force proportional to $V_{clamp} - V(t)$. 
In general the probability $p_o(t)$ is specified by a set of rate equations which reflect the dynamics of the channels, 
a description introduced by Hodgkin and Huxley in their famous 1952 paper \cite{hodgkin_quantitative_1952}. 
We come back to this point 
in Section III. For now we note that if we hold $V_{clamp}$ at a large negative value, a steady state solution 
of (\ref{eq: membrane}) exists such that $V(t) = V_r$ is also large and negative, and therefore $p_o \approx 0$ 
(see Fig. \ref{p_open}), namely: 

\begin{equation}
V_r = \frac{N_0 \chi_{\ell} V_N + V_{clamp} / R}{N_0 \chi_{\ell} + 1 / R} 
\label{eq: resting_pot}
\end{equation}

\begin{figure}
	\includegraphics[width=\linewidth]{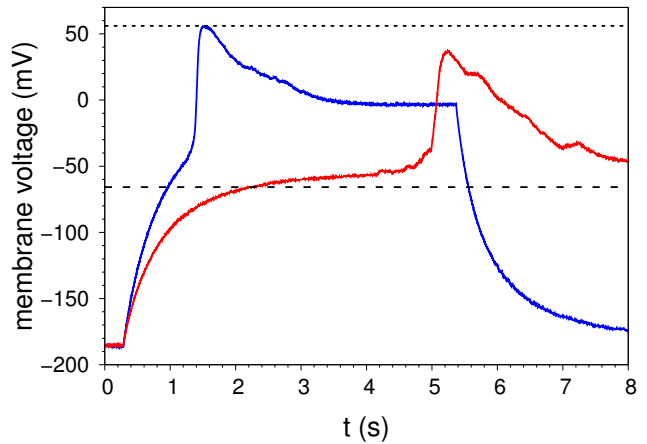}
	\caption{Time course of two action potentials recorded in the physical AA, demonstrating the delay in firing 
approaching threshold. For both traces, the CLVC was held at $V_{clamp} = - 200 \, mV$ before being stepped up 
at $t = 0.28 \, s$.  The clamp was raised to $0$ for the blue (short delay) trace and to $- 54.55 \, mV$ 
for the red (long delay) trace. 
The thick dashed line shows the threshold $V_{crit} \approx - 66 \, mV$, calculated as explained 
in the text, and the thin dotted line shows the Nernst potential $V_N$. The break in the blue trace at $t \approx 5.5 \, s$ 
corresponds to the CLVC being stepped down again to $V_{clamp} = - 200 \, mV$.}
	\label{v_trace_exp}
\end{figure}

\noindent From this expression we see that the CLVC fills the role of a second ionic gradient in determining the resting 
potential, with $V_{clamp}$ corresponding to the Nernst potential for this second species (e.g. $\mbox{Na}^+$) and 
$1/R$ corresponding to the leak conductance times the number of channels. The CLVC resistance $R$ is chosen 
such that the clamp current is sufficient to pull the resting potential to negative values $V_r \sim - 100 \, mV$ 
with channels closed, 
and on the other hand, such that the channel current with channels open can overwhelm the clamp current 
so that the AA can fire. From (\ref{eq: membrane}) and (\ref{eq: resting_pot}) one finds that we require 
$N_0 \chi_{\ell} < 1 / R << N_0 \chi$. Typical values of the parameters for the AA with the KvAP channels 
are: $C \approx 300 \, pF$, $N_0 \sim 100$, $\chi \approx \cfrac{10 \, pA}{50 \, mV} 
= 2 \times 10^{- 10} \, \Omega^{-1}$, $\chi_{\ell} / \chi \sim 1 / 100$, $R \sim 10^9 \, \Omega$, 
$V_N \approx + 50 \, mV$, $V_{clamp} \approx - 120 \, mV$. As a result, $V_r \approx - 90 \, mV$, 
currents are in the $100 \, pA - 1 \, nA$ range, and the characteristic time scale $C / (N_0 \chi) \sim 10 \, ms$ 
while $C / (N_0 \chi_{\ell}) \sim 1 \, s$. Under these conditions, starting from the resting potential 
$V(t) = V_r$, a perturbation such as a positive input current (which, in the experiment, can be delivered 
through a separate current clamp) can cause the system to fire an action potential \cite{Hector2017}. 
Referring to Fig. \ref{v_trace_exp}, the rising edge is caused by opening of the potassium channels, 
and has a universal shape; the falling tail is caused by the channels going into the inactive state, which allows 
the CLVC to pull the voltage down. Elsewhere we will discuss the requirements on the channel dynamics 
in order for the system to exhibit self-sustained oscillations (that is, a firing rate) in the presence of a steady 
input current; here we focus on the threshold for firing, which is a critical point of the corresponding dynamical 
system. 

\section{Experiment}

\noindent We have seen in Section I that the AA is a much simplified version of a spiking neuron. Nonetheless, there are 
aspects of the dynamics of the neuron which possess a degree of universality, so we expect them to remain 
the same independent of the details of the system. In particular, the behavior near a critical point, i.e. the threshold 
for firing. To explore the neighborhood of this critical point, we use as control parameter the command voltage to 
the CLVC ($V_{clamp}$). Physically this differs slightly from traditional electrophysiology, as the CLVC 
is neither a voltage clamp nor a current clamp, but rather something in between. In a neuron where the action potential 
is shaped by sodium and potassium currents, the analogous control parameter would be the Nernst potential of 
the potassium ions. The experimental protocol is as follows. We start from a steady state with $V_{clamp}$ 
held at $- 200 \, mV$, which typically corresponds to $V_r \approx - 180 \, mV$.  At these voltages the channels are fully closed, but not inactivated. The clamp command voltage $V_{clamp}$ is then stepped up to various values in the range 
between $0$ to $- 100 \, mV$, and held steady. The consequent increase in the AA voltage $V(t)$
is analogous to charging an RC circuit, with the membrane acting as the capacitor and the leak current and clamp resistor acting as the resistances. If functional ion channels are in the membrane, and $V_{clamp}$ is above a threshold 
$V_{crit}$, the AA will fire, i.e. a voltage spike will occur. The current through the membrane due to the channels opening is much larger than the CLVC can provide, which  causes the voltage to spike towards the Nernst potential. 
If $V_{clamp}$ is below $V_{crit}$, the AA does not fire and voltage will stabilize at $V(t) \approx V_{clamp}$. After the AA has fired, $V_{clamp}$ is stepped back to $- 200 \, mV$ and held there for 
$\sim 20s$ to allow the channels to recover from inactivation; then the system is probed again 
through another cycle. One challenge in these experiments is the stability in terms of maintaining the same threshold 
over a set of measurements meant to explore the vicinity of the critical point. Thresholds may change over time 
due to changes in the number of channels in the lipid bilayer, and changes in the capacitance (due to expansion 
or shrinking of the lipid bilayer). \\

\subsection{Firing Time Delay}
When $V_{clamp}$ is stepped above threshold, the KvAP channels will open to initiate a spike and then inactivate, preventing further spikes until they are allowed to recover at resting potentials $< - 120 \, mV$. The spikes that occur above threshold are qualitatively similar regardless of $V_{clamp}$, a key feature that defines action potentials.  Instead, the value of $V_{clamp}$ determines the amount of time that elapses before the onset of the action potential.  Namely, we find that as the membrane voltage is stepped closer to the voltage threshold $V_{crit}$ from above, there is a significant delay in the time it takes to depolarize the inner chamber. The measure of delay $\tau$ we adopt here is the time interval between the step in $V_{clamp}$ and the peak of the action potential. This differs slightly from the time of spike onset, with the main benefit being the ease of locating the peak versus the threshold. In any case, both methods produce 
equivalent results in what follows.  We have also explored alternative definitions, such as measuring the first derivative peak, with similar results for the scaling of $\tau$. \\

\begin{figure}
	\includegraphics[width=\linewidth]{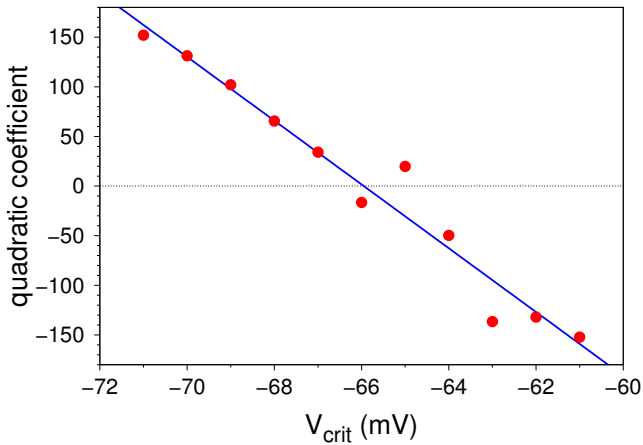}
	\caption{Graph of the quadratic coefficients - the curvature, basically - of log-log plots 
	of $\tau$ vs $(V_{clamp} - V_{crit})$, for different values of $V_{crit}$, as explained in the text. 
	The x-intercept of the linear fit (blue line) is taken as the actual threshold value $V_{crit}$ for the purpose 
	of obtaining scaling exponents. In this case we find $V_{crit} \approx - 66.0 \, mV$. }
	\label{fig:quadratic_exp}
\end{figure}

\subsection{Determining the Threshold}

\noindent We expect the delay time $\tau$ to scale with the distance to the critical point $(V_{clamp} - V_{crit})$, 
as explained in Section III. To experimentally confirm this relation, we must first locate $V_{crit}$ with precision.  Unfortunately, there is no uniform method to locate this threshold and it is generally accepted to be an empirical quantity \cite{platkiewicz_threshold_2010}, though some models do attempt to calculate it numerically \cite{angelino_excitability_2007}.  We take a slightly different approach given the fact that $V_{crit}$ is the critical point of our system. To find $V_{crit}$, we assume the existence of this critical point and proceed as follows. 
We take a series of action potentials obtained from the same AA prep and measure the delay time $\tau$.  The delay is subsequently plotted against $(V_{clamp} - V_{crit})$ on a log-log plot, with an initial estimate for $V_{crit}$ being the empirically observed value during experiment. We then adjust $V_{crit}$ to make the points fall on a straight line. 
This is a usual procedure in the study of critical phenomena, and we find that it gives a more precise 
- and different - determination 
of the critical point compared to other methods used in electrophysiology, which we summarize later. In detail, 
we produce, for the same data set, log-log plots of $\tau$ vs $(V_{clamp} - V_{crit})$, for different values of 
$V_{crit}$. We fit each plot to a straight line, subtract the value of the fit from the experimental data, and 
fit a quadratic form to these residuals. We then determine the value of $V_{crit}$ which corresponds to the 
coefficient of the quadratic term vanishing (Fig. \ref{fig:quadratic_exp}). \\

\begin{figure}
	\includegraphics[width=\linewidth]{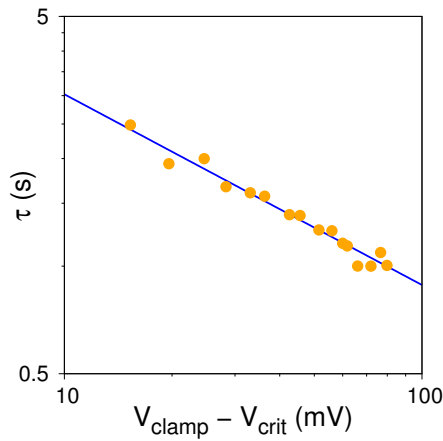}
	\caption{Log-log plot of delay time $\tau$ versus distance to threshold $(V_{clamp}- V_{crit})$, obtained from 
	one set of recordings in the physical AA. This plot is used to determine the critical exponent. 
	The threshold used for this data set, determined using the method of Fig. \ref{fig:quadratic_exp}, 
	is $V_{crit} = - 88.6 \, mV$. Fitting with a straight line returns the exponent 
	-0.53, close to the value of -1/2 expected for the saddle-node bifurcation. }
	\label{fig:loglog_22520}
\end{figure}

\begin{figure}
	\includegraphics[width=\linewidth]{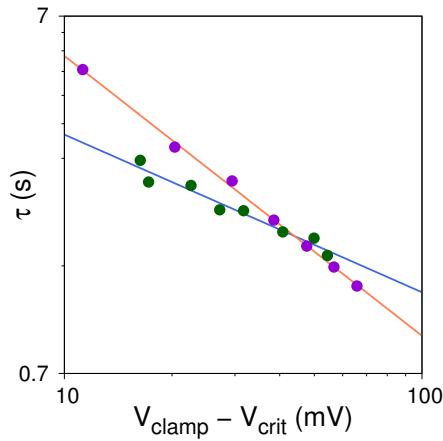}
	\caption{Two further data sets relating $\tau$ to $(V_{clamp} - V_{crit})$ in the physical AA.  The fits return 
	exponents of -0.44 and -0.77 for the green and purple dots, respectively. Reasons for the deviation from 
	the predicted value are discussed in the text.}
	\label{fig:loglog_other2}
\end{figure}

If we want to measure critical exponents, a precise 
determination of the threshold is essential, where by threshold we mean the location of the critical point 
in parameter space. The methods that have been employed in electrophysiology for estimating action potential thresholds were also considered. The well known methods all relate to observing changes in the 
derivative of the voltage \cite{sekerli_estimating_2004}.  From our perspective, these methods have a tendency to overestimate the voltage threshold, due to the fact that channels have a slight delay in opening and by the time 
there is a significant peak in the derivative the voltage has already risen beyond the threshold. Indeed, the location 
of the threshold shown in Fig. \ref{v_trace_exp} is {\it not} where one would expect from a qualitative examination 
of the time traces. It does however correspond to the location of the critical point in parameter space, as we show 
in Section III. For similar recordings and threshold determination in real neurons, see for instance 
\cite{Wickens1988, Kress2008}.  \\
\noindent The result of the scaling assumption for one data set is shown in Fig. \ref{fig:loglog_22520}. The slope of the 
fitted straight line is $- 0.53$, close to $- 1/2$. Unfortunately the range of the measurements is quite limited, 
covering only a decade in the control parameter $(V_{clamp} - V_{crit})$. As we will see in Section III, 
the relatively small number of channels in the AA limits how close one is able to get to the critical point, 
and beyond that, the stability and noise in the system. Fig. \ref{fig:loglog_other2} shows plots for two more data sets 
(independent preps of the AA).  These data sets show differing values of the scaling exponent, we suspect due to the drifting of the threshold during the experiment as mentioned earlier.  In particular, excess channels that do not initially insert into the membrane may slowly insert over time.  This causes the number of channels in the membrane to increase over time which alters the threshold (see section III and Mat. \& Met.). \\

\subsection{Materials and Methods}
\noindent The artificial axon (AA) system consists of voltage gated ion channels from the Archaea \textit{Aeropyrum Pernix} (KvAP) inserted into a phospholipid bilayer (black lipid membrane).  The membrane rests on a $\sim$200 $\mu$m aperture at the bottom of a plastic centrifuge cup (Beckman-Coulter), which is firmly held in a custom made Teflon chamber. The lipid membrane separates the Teflon compartment and the inside of the plastic cup. The outer chamber is grounded; the inner chamber contains two 1 mm diameter pellet AgCl electrodes (Warner Instruments E205), one connected to the CLVC and one to measure the voltage $V(t)$. Data is recorded using a custom LabView 2014 program. Schematics for the head stage and clamp amplifiers can be found in \cite{Amila2016}.

\begin{figure}
	\includegraphics[width=\linewidth]{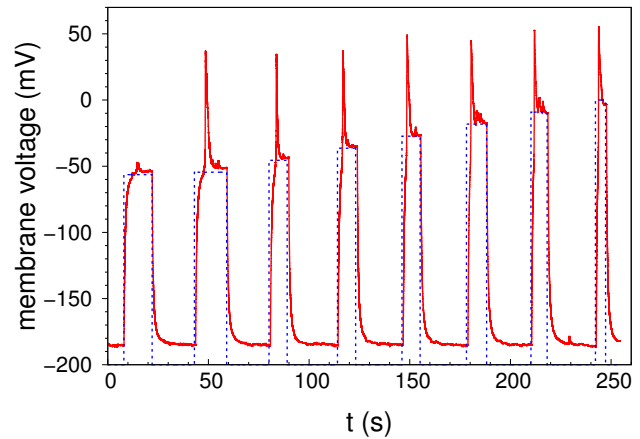}
	\caption{Series of action potentials evoked by stepping the CLVC (dashed line showing $V_{clamp}$) 
	to different values.  For the first cycle $V_{clamp}$ is set below the threshold for firing $V_{crit}$.}
	\label{fig:v_trace_full}
\end{figure}

The procedure to build an Artificial Axon (AA) prep is as follows: 
DPhPC (1,2-diphytanoyl-sn-glycero-3-phosphocholine dissolved in chloroform at 25 mg/mL (Avanti Polar Lipids) is Nitrogen dried and redissolved in 12.5 $\mu$L decane for a concentration of 20 mg/mL.  A small droplet of the decane-lipid mixture is dripped via syringe onto the aperture of the aforementioned plastic cup, which serves as the intracellular chamber 
(Fig. \ref{fig:setup}).  The cup is placed under vacuum for 30 minutes to ensure that the chloroform is thoroughly removed before the cup is inserted into the Teflon chamber. The outer chamber is filled with 150 mM KCl and 10 mM Hepes  pH 7.0 while the plastic cup is filled with 30 mM KCl and 10 mM HEPES plus 120 mM sucrose pH 7.0 to equalize osmotic pressure. 
Additional DPhPC is pipetted in at the bottom of the cup to seal the opening and prevent the diffusion of ions.  At this point the electronics are turned on and the CLVC injects current into the inner cup which clamps the voltage at the desired value.  All experiments start with the voltage clamped at -200 mV after the system is prepared. A planar lipid bilayer is then painted into the opening and allowed to stabilize.
Expressed KvAP solubilized in DPhPC vesicles are then removed from an $-80 ^{\circ}$C freezer and thawed quickly at room temperature.  
This solution is mixed with a pipette prior to a small amount (.3-1.5 $\mu$L) being injected close to the lipid bilayer.  The channels are given a few minutes to insert, then the CLVC is stepped up from its initial value of -200 mV to confirm the insertion of channels.  The occurrence of an action potential signifies that a sufficient number of channels have inserted into the membrane.  A typical voltage trace from which the measurements 
are obtained is displayed in Fig. \ref{fig:v_trace_full}.  

The ion channels themselves were expressed in \textit{E coli.} following published procedures \cite{Amila2016}, which were adapted from \cite{ruta_functional_2003}.  The procedure for reconstitution of the channels into lipid vesicles also follow that of \cite{Amila2016}, with the adjustment of the lipid to protein ratio to 3 instead of 1 (wt/wt).

\section{Theory} 

\noindent Models of the neuron as a dynamical system \cite{Koch_Book} hail back to the ground breaking studies 
of Hodgkin and Huxley \cite{hodgkin_quantitative_1952} and form some of the most quantitative descriptions 
in neurobiology. The AA is a much simplified version of a neuron, a minimal system on which 
some of the universal characteristics of the more complex living matter can be illustrated and studied. 
Such is the threshold for firing an action potential, which is one of the most important and defining characteristics 
of neurons. 
It is not difficult to see that, in the AA, the threshold for firing corresponds to the critical point of a saddle node 
bifurcation. We show this in subsection B, after briefly recalling the general theory \cite{Strogatz_Book} in subsection A. 
Then we discuss critical behavior in the simplest model with channel dynamics similar to the AA (subsection C). 

\subsection{Normal form of the saddle-node bifurcation}

\noindent Consider the 1D dynamical system: 

\begin{equation}
\dot x  = r  + x^2
\label{eq: norm_form}
\end{equation}

\noindent where $r$ is a control parameter. From the graph of $\dot x$ vs $x$ (a parabola) we see that for $r < 0$ 
the system has two fixed points ($\dot x = 0$), one stable and one unstable. They merge at the critical point 
$r = 0$; for $r > 0$ the velocity is always positive so the system escapes to infinity. However, near the critical point 
it exhibits critical slowing down: for $r > 0$ (\ref{eq: norm_form}) integrates to: 

\begin{equation}
x(t)  = \sqrt{r} \, tan(\sqrt{r} \, t + b) \quad , \quad b = arctan(x(0) / \sqrt{r})
\label{eq: norm_form_2}
\end{equation}

\noindent We see from (\ref{eq: norm_form_2}) that the time to escape to infinity is finite; starting from a 
large negative value $x(0)$, so that $b \approx - \pi / 2$, that time is: 

\begin{equation}
\tau  \sim  \frac{\pi}{\sqrt{r}} \, \propto r^{- 1/2}
\label{eq: norm_form_3}
\end{equation}

\noindent which can also be estimated by: 

\begin{equation}
\tau  \sim  \int_{- \infty}^{+ \infty} \frac{d x}{r + x^2} \, = \, \frac{\pi}{\sqrt{r}} \, = \, \pi \, (r - r_c)^{- 1/2}
\label{eq: norm_form_4}
\end{equation}

\noindent The delay time diverges as one approaches the critical point from above, and scales with the 
distance to the critical point with characteristic exponent $- 1/2$.

\subsection{Saddle-node bifurcation in the Artificial Axon}

\noindent The arguments in this section closely follow those developed in \cite{Morris_Lecar}. Consider 
the "membrane equation" (\ref{eq: membrane}) for the AA. To show the existence of the bifurcation, we make 
the approximation of considering the channel dynamics fast enough that the probability that channels are 
open $p_o(t)$ in (\ref{eq: membrane}) always is the equilibrium value $p_o(V)$ shown in 
Fig. \ref{p_open}, and we ignore the inactive state. The exact meaning of this approximation is seen 
more clearly in the next section. In this case, (\ref{eq: membrane}) becomes the 1D dynamical system: 

\begin{equation}
\frac{d V}{d t} = - \frac{N_0 \chi}{C} \, [ p_o(V) + \chi_{\ell} / \chi ] [V(t) - V_N] -  
\frac{1}{RC} \, [V(t) - V_{clamp}]
\label{eq: membrane2}
\end{equation}

\noindent where $p_o(V)$ is a Fermi - Dirac function: 

\begin{equation}
p_o(V) =  \frac{1}{e^{-q (V - V_0) / k T} + 1} 
\label{eq: Fermi_Dirac}
\end{equation}

\noindent with given values of the parameters $q$ and $V_0$ ($k T$ is the thermal energy), as shown 
in Fig. \ref{p_open}. The fixed points ($dV / dt = 0$) of the dynamical system (\ref{eq: membrane2}), 
(\ref{eq: Fermi_Dirac}) are given by:  

\begin{equation}
p_o(V) =  \frac{1}{N_0 \chi R} \, \left ( \frac{V - V_{clamp}}{V_N - V} \right ) \, - \, \frac{\chi_{\ell}}{\chi}
\label{eq: crit_point_1}
\end{equation}

\begin{figure}[h]
	\centering
	\subfigure[]{
		\includegraphics[width= \linewidth]{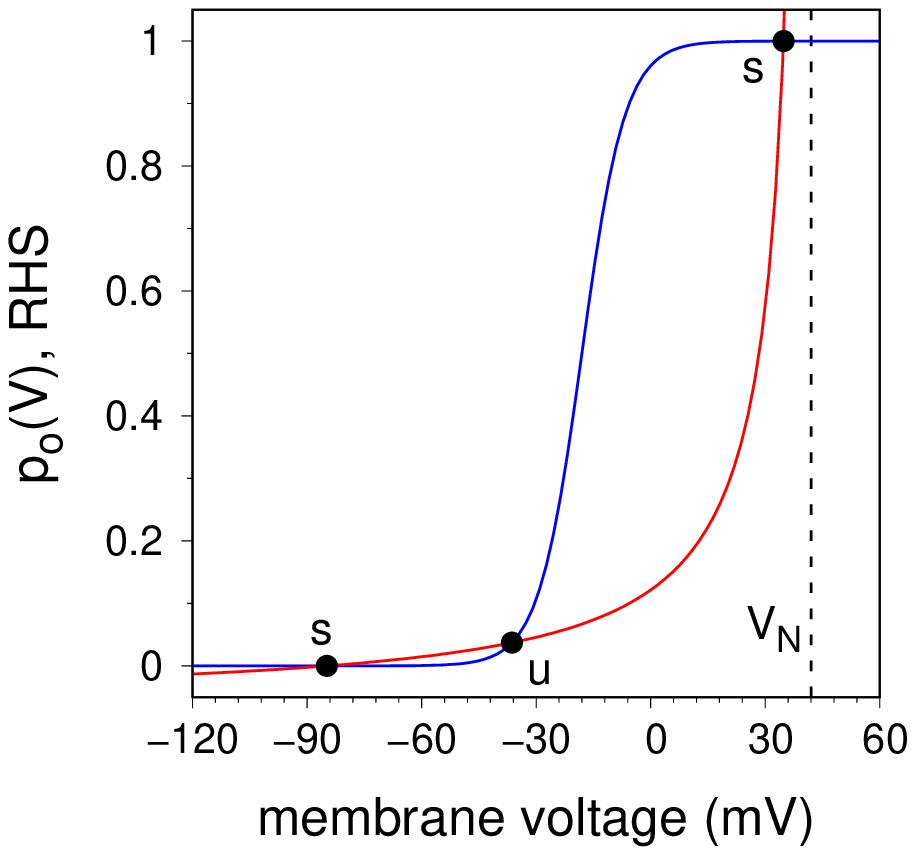} }
	\centering
	\subfigure[]{
		\includegraphics[width= \linewidth]{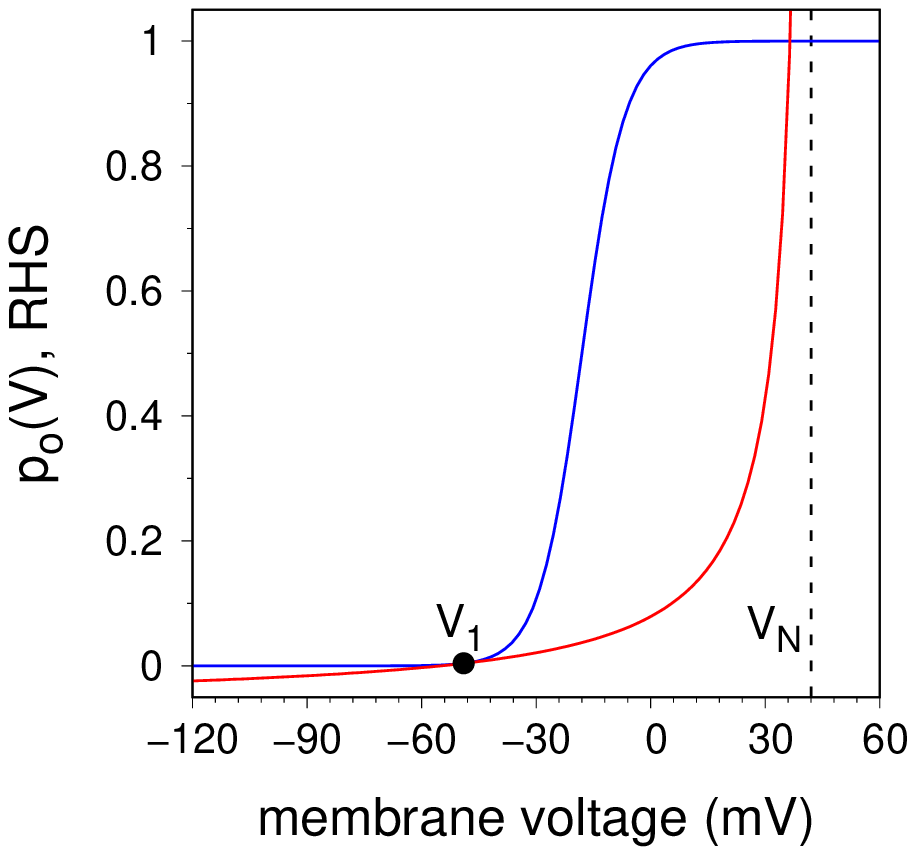} }
\caption{Plots of the LHS and the RHS of equation (\ref{eq: crit_point_1}); the dotted asymptote 
indicates the Nernst potential $V_N$. \\
a) for $V_{clamp} < V_{crit}$ there are 3 fixed points, marked stable (s) or unstable (u). \\
b) Same plot as in( a), for $V_{clamp} = V_{crit}$, showing the location of the critical point $V_1$.}
	\label{fig:bifurcation}
\end{figure}

\noindent Fig. \ref{fig:bifurcation} shows plots of the LHS and the RHS of (\ref{eq: crit_point_1}), the relevant 
range of voltages being $V_{clamp} < V(t) < V_N$. 

\noindent We see that for $V_{clamp}$ sufficiently negative (Fig. \ref{fig:bifurcation}a) there are 3 fixed points. As $V_{clamp}$ is increased, the stable and unstable fixed 
points on the left merge at a critical point (Fig. \ref{fig:bifurcation}b) where the two curves in the figure are tangent 
to each other. For larger values of $V_{clamp}$ these two fixed points disappear and only the stable fixed point 
close to the Nernst potential remains. Evidently this is the same bifurcation as exhibited by the dynamical system 
(\ref{eq: norm_form}), so we expect the same behavior near the critical point. The disappearance of the lower stable fixed point for $V_{clamp} > V_{crit}$ corresponds to firing of the action potential. 
If $V_{clamp}$ is stepped close to, but above, $V_{crit}$, there will be a delay time for firing, scaling as: 

\begin{equation}
\tau \sim  (V_{clamp} - V_{crit})^{- 1/2}
\label{eq: delay_time}
\end{equation}

\noindent Explicitly, the critical point $V_{clamp} = V_{crit}$, $V(t) \equiv V_1$ is defined by: 

\begin{normalsize}
\begin{equation}
\begin{aligned}
\left\{\begin{array}{ll}
 p_o(V_1) = \cfrac{1}{N_0 \chi R}\left(c\frac{V_1-V_{crit}}{V_N-V_1}\right)-\cfrac{\chi_{\ell}}{\chi} \\
\\
\left. \cfrac{d p_o}{d V}\right \rvert_{V = V_1}  = \cfrac{\partial}{\partial V} \left (\cfrac{1}{N_0 \chi R} \, 
\left. \cfrac{V - V_{crit}}{V_N - V} \right ) \right \rvert_{V_1} \\
\\
= \cfrac{1}{N_0 \chi R} \, \cfrac{V_N - V_{crit}}{(V_N - V_1)^2}
\end{array}\right.
\end{aligned}
\label{eq: crit_point_condition}
\end{equation}
\end{normalsize} 

\noindent Set $V_{clamp} = V_{crit} + \epsilon$ and write (\ref{eq: membrane2}) in the form: 

\begin{equation}
\frac{d V}{d t} = F(V, V_{clamp})
\label{eq: membrane_3}
\end{equation}

\noindent Expanding around the critical point: 

\begin{equation}
F(V, V_{crit} + \epsilon) \approx F(V, V_{crit}) + \frac{1}{R C} \, \epsilon 
\label{eq: crit_point_2}
\end{equation}

\noindent while 

\begin{equation}
\begin{split}
F(V, V_{crit}) \approx F(V_1, V_{crit}) + \frac{\partial}{\partial V} F(V, V_{crit}) \vert_{V_1} \, (V - V_1) \\
\, + \,  \frac{1}{2} \, \left. \frac{\partial^2 F}{\partial V^2} \right \vert_{V_1} \, (V - V_1)^2 
\end{split}
\label{eq: crit_point_3}
\end{equation}

\noindent Using (\ref{eq: crit_point_1}) one recognizes that the first two terms on the RHS of (\ref{eq: crit_point_3}) 
vanish. Also, the coefficient of the quadratic term is positive. Finally, 

\begin{equation}
\begin{split}
F(V, V_{crit} + \epsilon) \approx \frac{1}{R C} \, \epsilon + b \, (V - V_1)^2 \\
\\
b = \frac{1}{2} \, \left. \frac{\partial^2 F}{\partial V^2} \right \vert_{V_1} > 0 \quad , \quad 
\epsilon = V_{clamp} - V_{crit} > 0 
\end{split}
\label{eq: crit_point_4}
\end{equation} 

\noindent Thus, close to the critical point the dynamical system (\ref{eq: membrane2}) reduces to 
(\ref{eq: norm_form}). Using the estimate (\ref{eq: norm_form_4}) for the delay time: 

\begin{equation}
\begin{split}
\tau \sim \int_{- \infty}^{+ \infty} \frac{d V}{\epsilon / RC + b (V - V_1)^2} \\ 
\\
 = \frac{\pi}{\sqrt{b \epsilon / (R C)}} \, = \, \pi \, \sqrt{\frac{R C}{b}} \,  (V_{clamp} - V_{crit})^{- 1/2} 
\end{split}
\label{eq: crit_point_5}
\end{equation} 

\noindent and, noting that $b \propto N_0 \chi / C$: 

\begin{equation}
\tau \propto \frac{R C}{\sqrt{N_0 R \chi}}  \,  (V_{clamp} - V_{crit})^{- 1/2} 
\label{eq: crit_point_6}
\end{equation} 

\begin{figure}[h]
	\includegraphics[width=\linewidth]{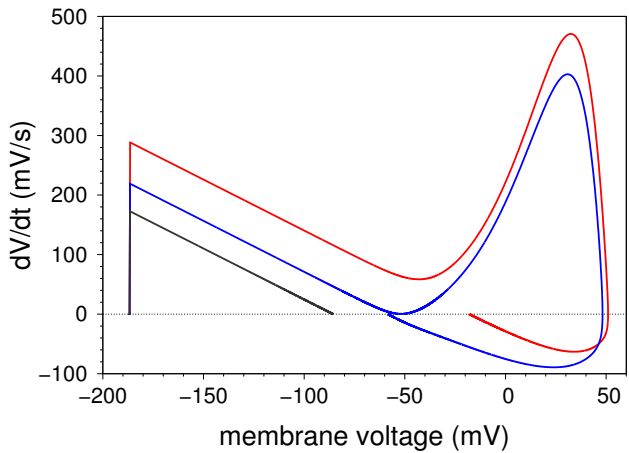}
	\caption{Phase space visualization of three different trajectories obtained from integrating Model A 
	of Section III C. $V_{clamp}$ was stepped to $-86 \, mV$ for the black curve, and being below threshold, failed to 
	generate an action potential.  For the blue and red trajectories the CLVC was stepped to 
	$-58.8$ and $-18.18 \, mV$ respectively.  The blue trajectory, reaching closer to the saddle-node remnant,  
	experiences a significant delay in firing compared to the red trajectory.}
	\label{phaseplot}
\end{figure}

\noindent The delay time scales with the distance to the critical point with the exponent $- 1/2$, with a prefactor 
proportional to the characteristic $RC$ time scale, modulated by a factor which depends on the number of channels 
and the ratio between the open channel conductance and the CLVC conductance. \\ 
\noindent The rate at which the voltage proceeds to the fixed point close to the Nernst potential can be visualized in a phase space plot. Fig. \ref{phaseplot} shows a plot of $dV / dt$ vs $V$ for several trajectories (time traces) obtained 
from integrating the AA model described in the next section. When $V_{clamp}$ is stepped up, $dV / dt$ jumps 
to a positive value and the voltage immediately starts to rise. As $V$ approaches the critical point $V_1$, $dV / dt$ 
will reach its minimum value, which can be arbitrarily small depending on how close $V_{clamp}$ is to $V_{crit}$. The scaling of the delay time $\tau$ can be attributed to this bottleneck region.  Eventually, $dV/dt$ will start to rise again and spiking will occur, as shown in the right half of Fig. \ref{phaseplot}. The subsequent drop and sign change in velocity is due to the inactivation of the channels which moves the fixed point back near $V_{crit}$.

\noindent The theory in this section is not new; specifically, this dynamical systems approach appears in the 
excellent article by Morris and Lecar \cite{Morris_Lecar}, where they describe their electrophysiology measurements 
on the barnacle muscle fiber. We have presented the calculations in detail to show the generality as it pertains 
to the AA. In summary: for the delay time $\tau$ measured in the experiments of Section II, we expect a scaling 
exponent of $- 1/2$. However, (\ref{eq: crit_point_6}) also shows that in order to see that scaling, the experiment must 
be stable with respect to the parameters appearing in the prefactor to the power law. If, during the course of one 
run, the capacitance $C$ changes (because the membrane shrinks), 
or the number of channels (and thus the parameter $b$) 
changes (for example, some channels may diffuse to the rim of the membrane and effectively disappear; new 
channels may insert), the exponent measured from the log-log plot of the data may be affected. We believe such a drift 
affected the data sets in Fig. \ref{fig:loglog_other2}, which show an exponent 
significantly different from $-1/2$ (discussed in section IV).

\subsection{Model with channel dynamics} 

\begin{figure}[h]
	\centering
	\subfigure[]{
		\includegraphics[width= .4\linewidth]{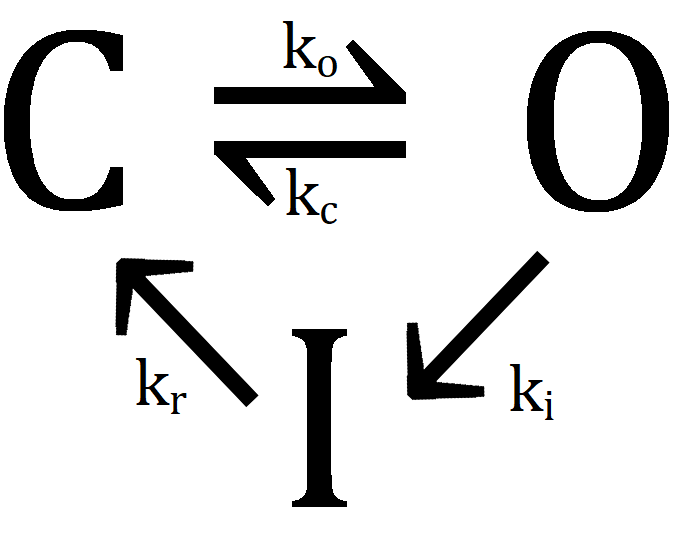} }
	\hspace{.1\linewidth}
	\subfigure[]{
		\includegraphics[width= .4\linewidth]{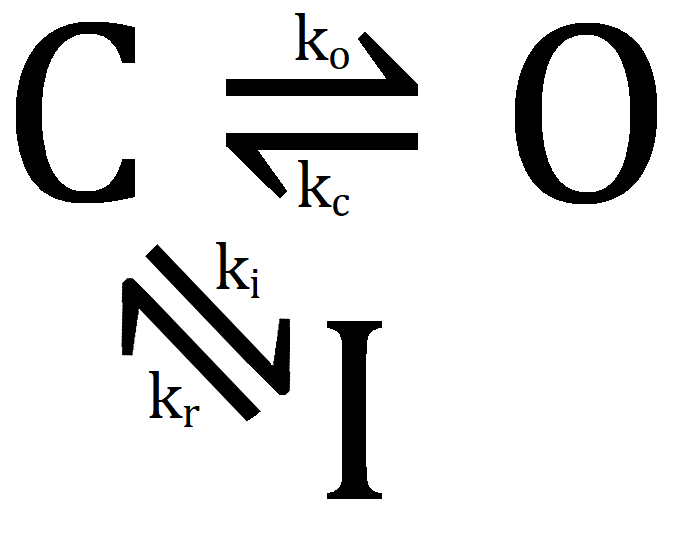} }
	\caption{Two minimal models for the opening dynamics of the KvAP channel. C: closed state; 
O: open state; I: inactive state. The rates are voltage dependent (see text). \\}
	\label{fig:models}
\end{figure}

\noindent In order to represent the complete action potential seen in Fig. \ref{v_trace_exp} - not just its rising front - 
we need to introduce channel dynamics in the AA model, specifically the inactive state. The minimal model 
for the KvAP channel has 3 states: closed, open, and inactive (which is functionally closed), and voltage dependent 
rates to move between these states. The actual dynamics of the channels - if one looks in detail - is significantly 
more complex \cite{schmidt_gating_2009}, with more states and rates. For instance, the KvAP is a tetramer, and the 
conformational state of each subunit must in principle be specified for a more complete description of the state 
of the channel. However, these microscopic details are often not important for the description of the kind of 
collective phenomena - such as action potentials - that we want to explore with the AA. In addition, 
a more complete description leads to the introduction of several more rates, i.e. an uncomfortable proliferation of the number 
of parameters in the model. We therefore start from a minimal model of the channels which still contains 
the important dynamics: it is depicted schematically in Fig. \ref{fig:models}(a), where C, O, and I represent the closed, open, and inactive states, respectively. 
The main feature of the model (which we will call 
"Model A") is that the inactive state is accessed from the open state and flows into the closed state. The 
unidirectional arrows are strictly speaking unphysical (the equilibrium state of this system, in the sense of a state 
where the probabilities of the three states do not change in time, violates detailed balance), but they represent 
a permissible approximation if the rates for the transitions $C \rightarrow I$ and $I \rightarrow O$ are small. 
The key advantage obviously being that we only have four rates. Without introducing more rates, the only other possible model 
compatible with the known overall KvAP dynamics is Model B shown in Fig. \ref{fig:models}b. It is not immediately clear 
which of these two models is a better representation for the KvAP; for our discussion of scaling near the 
critical point they lead to the same result for universal quantities such as the scaling exponent, but other important 
behavior may differ. We also note that if the time scales $1 / k_o$, 
$1 / k_c$ are much faster than any other time scale in the system (this is {\it not} the case for the 
present AA), then the dynamics of the two models are equivalent. In the following, we will stay 
with Model A. \\ 
Since the channels are voltage gated, the rates $k_o$, $k_c$ must be made voltage dependent. In the usual way, 
we assume a 1D barrier crossing process and write: 

\begin{equation}
k_o(V) = \kappa \, e^{\alpha (V - V_0)} \quad , \quad k_c(V) = \kappa \, e^{- \alpha (V - V_0)}
\label{eq: rates_1}
\end{equation} 

\noindent This "symmetric" form contains the extra assumption that the transition state lies exactly "in the middle" 
between the open and closed states along the 1D conformational landscape; it can be relaxed at the cost 
of introducing one more parameter. The rates (\ref{eq: rates_1}) must be related to the equilibrium 
probability $p_o(V)$ of Fig. \ref{p_open} and equation (\ref{eq: Fermi_Dirac}) by $k_o / (k_o + k_c) = p_o(V)$. 
Thus, in terms of the parameters in (\ref{eq: Fermi_Dirac}), $\alpha = - q / (2 kT)$. Similarly, for the rates 
pertaining to the inactive state we write: 

\begin{equation}
k_i(V) = \kappa_i \, e^{\alpha_i (V-V_{0}^i)} \quad , \quad k_r(V) = \kappa_r \, e^{- \alpha_r (V-V_{0}^r)}
\label{eq: rates_2}
\end{equation} 

\noindent We use values for the 
parameters which are based off of the experimental data, by fitting the model below onto the voltage traces 
in the experiment; they are: \\ \\
$\kappa = 0.3 \, s^{-1}$, $\alpha = 0.0887 \, (mV)^{-1}$, $V_0 = - 18$ mV, $\kappa_i = 0.01 \, s^{-1}$, 
$\alpha_i = -0.1 \, (mV)^{-1}$, $V_{0}^i = -80 \, mV$, $\kappa_r = 1.3 \, s^{-1}$, $\alpha_r = 0.02 \, (mV)^{-1}$, 
$V_{0}^r = -50 \, mV$.  \\ \\
The AA model which includes channel dynamics (according to Model A) consists of equation (\ref{eq: membrane}), 
which we re-write here: 

\begin{equation}
\frac{d V}{d t} = - \frac{N_0 \chi}{C} \, [ p_o(t) + \chi_{\ell} / \chi ] [V(t) - V_N] -  
\frac{1}{RC} \, [V(t) - V_{clamp}]
\label{eq: membrane_2}
\end{equation}

\noindent and the following rate equations for the probability that the channels are in the open state ($p_o(t)$), 
closed state ($p_c(t)$), or inactive state ($p_i(t)$; see Fig. \ref{fig:models}): 

{\renewcommand{\arraystretch}{2}
\begin{normalsize}
\begin{equation}
\begin{aligned}
\left\{\begin{array}{ll}
 \cfrac{d p_i}{d t} = p_o(t) k_i(V) - p_i(t) k_r(V) \\
 \cfrac{d p_o}{d t} = (1 - p_o - p_i) k_o - p_o (k_c + k_i) 
\end{array}\right.
\end{aligned}
\label{eq:rates_3}
\end{equation}
\end{normalsize}
}

\begin{figure}
	\includegraphics[width=\linewidth]{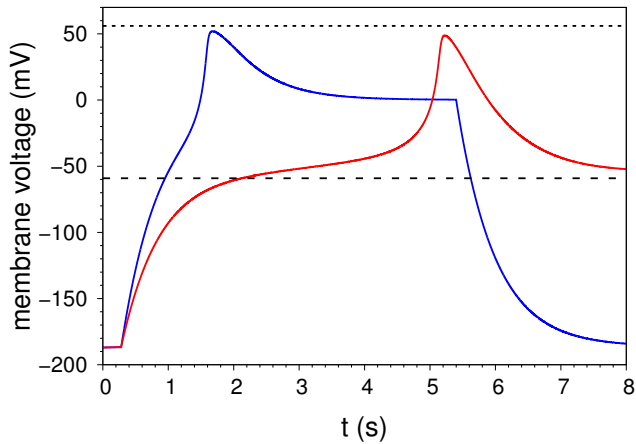}
	\caption{Action potentials produced from integrating the model in Section III C.  The clamp protocol and other settings follow those of the experimental results in Fig. \ref{v_trace_exp}. Namely, $V_{clamp} = - 200 \, mV$ initially;  
at $t = 0.28 \, s$ the clamp is raised to $0$ and $ -54.55 \, mV$ for the blue and red traces, respectively.  
The dashed line shows the threshold $V_{crit} \approx -59.1 \, mV$, determined as explained in the text; 
the dotted line shows the Nernst potential.}  
	\label{v_trace_sim}
\end{figure}
\begin{figure}
	\includegraphics[width=\linewidth]{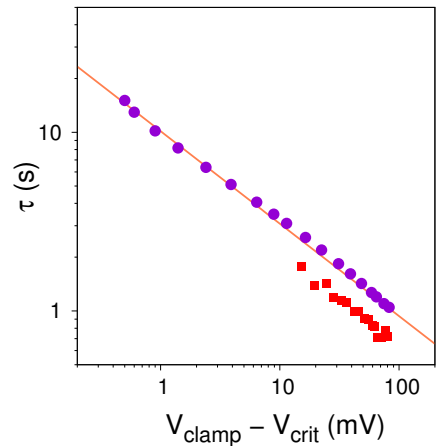}
	\caption{Log-log plot of the delay time $\tau$ vs distance to the critical point $(V_{clamp} - V_{crit})$ obtained 
from the simulation, using the same procedure used for the experimental data of Fig. \ref{fig:loglog_22520}. 
The threshold, determined as explained in the text, is $V_{crit} = -91.4 \, mV$ 
(for comparison, it is  $-88.6 \, mV$ for the experimental data plotted in Fig. \ref{fig:loglog_22520}).  
The linear fit returns a scaling exponent of -0.51, close to the experimental value of -0.53 
and the theoretical value of -1/2.  The experimental data of Fig. \ref{fig:loglog_22520} are shown as red squares 
for comparison; they are shifted downwards by $1.4 \, s$ for clarity. }
	\label{fig:loglog_sim}
\end{figure}


\noindent which enforce $p_o + p_c + p_i = 1$; the two equations (\ref{eq:rates_3}) are coupled to 
(\ref{eq: membrane_2}) through the voltage dependence of the rates (\ref{eq: rates_1}), (\ref{eq: rates_2}). \\ 
The 3D dynamical system (\ref{eq: membrane_2}), (\ref{eq:rates_3}) reproduces the phenomenology of the 
physical AA. Fig. \ref{v_trace_sim}, which is analogous to Fig. \ref{v_trace_exp}, shows two action potentials evoked in the model 
by stepping $V_{clamp}$ to different values.  We used this model to simulate the experiments 
of Section II, i.e. we produced a series of action potentials for different values of $V_{clamp}$, measured the delay 
time $\tau$, and determined the threshold $V_{crit}$ using the method described in Section II. 
Fig. \ref{fig:loglog_sim} shows the 
resulting log-log plot of $\tau$ vs $(V_{clamp} - V_{crit})$. We used the following parameter values in the model: \\ 
$C = 329.7$ pF, $\chi = 1.67 \times 10^{-10}$ $\Omega^{-1}$, $\chi_{\ell} / \chi = 8.8 \times 10^{-4}$, 
$R = 2 \times 10^9 \, \Omega$, $N_0 = 110$ , and the initial condition $V(t = 0) =  - 200 \, mV$ and 
$p_o(t=0) = p_i(t=0) = 0$ (i.e. channels closed initially). The advantage of the simulation 
is of course that we obtain a larger range of the power law behavior and thus a good determination 
of the scaling exponent; the linear fit in Fig. \ref{fig:loglog_sim} has a slope of $-0.51$. Here there is no question of the 
stability of the "experiment", since parameters such as $C$ and $N_0$ are fixed, but what limits how close 
one can get to the critical point (and thus how large a delay time $\tau$ one can observe) is the finite number 
of channels. \\

\section{Discussion}

\begin{figure}
	\includegraphics[width=\linewidth]{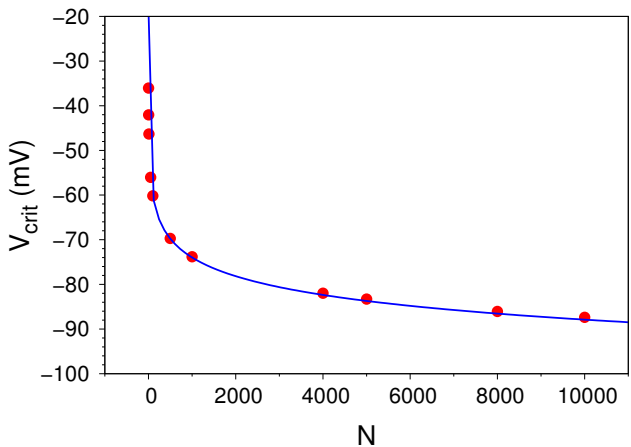}
	\caption{Numerically calculated relation between the number of channels in the system and the threshold, using steady state opening probabilities and no leak.  The blue fit is logarithmic with coefficient $-kT/q$ 
(see eq. (\ref{eq: Fermi_Dirac}) and \cite{platkiewicz_threshold_2010}).}
	\label{fig:NvsT}
\end{figure}

\begin{figure}
	\includegraphics[width=\linewidth]{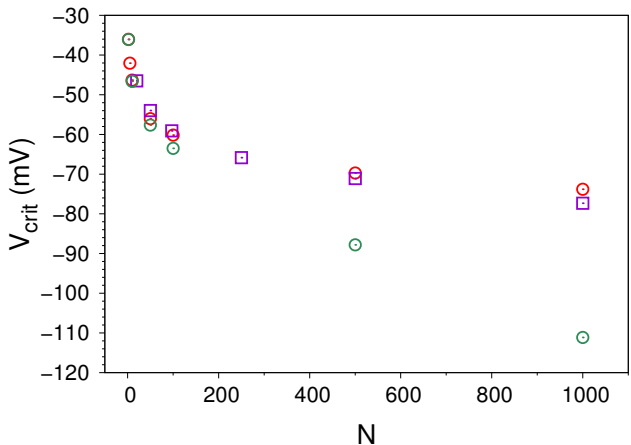}
	\caption{When a leak is introduced into the system, the threshold voltage is depressed.  The red circles are the same from Fig. \ref{fig:NvsT} ($\chi_{\ell}=0$), and the green circles are with $\chi_{\ell}/\chi = 10^{-3}$. The purple squares were simulated with model A and $\chi_{\ell}/\chi = 9.1 \times 10^{-5}$, they agree extremely well with the numerical results (not plotted).}
	\label{fig:NvsT_compare}
\end{figure}

Simulation of the model in section III allows us to explore regions close to the critical point that are difficult 
to access in the experiment.  Fig. \ref{fig:loglog_sim} shows that the simulation agrees well with experiment in the range where they overlap.  However, one limit still in effect is the finite size of the system, namely the number of channels.  Given that the number of channels in the simulation must match the experiment from which the rates were taken, the scaling exponent will only be correct up to a certain resolution, and we cannot achieve arbitrarily long delay times.  The simulations also demonstrate the 'stability' of the scaling exponent,  in that significantly varying the rates and conditions do not alter the scaling behavior correspondingly. This is of course to be expected since the scaling phenomenon arises from the saddle-node nature of the critical point.

Additional simulations of model A were performed to determine how the number of channels and the leak conductance affect the threshold.  Solutions to eq. (\ref{eq: crit_point_1}) with steady state opening were also calculated numerically to obtain a relation between $N_0$ and $V_{crit}$ for varying values of $\chi_{\ell}$. The results show that there exists a range of channels number $N_0$ such that excitability is possible, above and below which there is no bifurcation and thus no possibility of action potentials occurring. This fact can also be seen directly from the graphical representation 
of equation (\ref{eq: crit_point_1}) in Fig. \ref{fig:bifurcation} : increasing $N_0$ makes the slope of the RHS of 
(\ref{eq: crit_point_1}) shallower, and thus $V_{crit}$ 
more negative, but since $\chi_{\ell} / \chi > 0$ there is a maximum $N_0$ beyond which the only fixed point 
is the one close to $V_N$ (i.e. a steady state with channels open). Conversely, decreasing $N_0$  
makes the slope of the RHS of (\ref{eq: crit_point_1}) steeper, and thus $V_{crit}$ increases; 
the critical point $V_1$ (eq. (\ref{eq: crit_point_condition})) 
moves towards the inflection point of the open probability curve $p_o(V)$, and we see that if $N_0$ is too 
small, the unstable fixed point disappears, and so there is no firing. \\
In the case that there is no leak ($\chi_{\ell}=0$), we found a logarithmic relationship between the number of channels and the threshold.  This is in agreement with previous numerical studies \cite{platkiewicz_threshold_2010} calculating a relationship between the conductivity and the threshold voltage, since the total conductance is directly proportional to the number of channels.  For larger values of the leak, we find that $V_{crit}$ is significantly more negative and the relationship is no longer logarithmic.  When the leak is taken to be the same value as the simulation, the numerical solutions are in good agreement with the simulation of model A, indicating that the threshold does not depend strongly on the opening model.  These results are displayed in Figs. \ref{fig:NvsT} and \ref{fig:NvsT_compare}. \\
It is interesting to note that if the data from Fig. \ref{fig:loglog_other2} are taken and plotted with the simulated threshold rather than the fitted threshold, a scaling exponent much closer to the theoretical one is obtained.  This further supports our reasoning for the deviation in experimental behavior (i.e. the drift in voltage threshold as the experiment progressed). We expect that if a drift had occurred, it would have decreased the voltage threshold over time, as channels floating above the lipid membrane would have inserted over time, increasing the number of channels in the system. As such, a simulation using fits on the first few data points would be well suited to provide an estimate for how the system would behave in the absence of this drift.  Our simulation rates are fitted onto the spikes closest to the critical point, which were the data points first recorded in the experiment. \\
In conclusion, we report measurements of critical behavior in the AA, and corresponding simulations of a minimal 
model which describes the system well. Critical properties are universal and are thus the same for the AA, 
the model, and real neurons. Our quest is to construct, at each scale, the minimal system which retains 
the essential properties of the more complex biological systems of today. In this (primordial) world, the 
AA is the minimal neuron, networks of artificial axons may be developed into minimal nervous systems. 
More prosaically, the present work is an example of how the AA can be developed into a cell-free breadboard 
for electrophysiology research. \\

\begin{acknowledgments}
\noindent This work was supported by NSF grant DMR - 1809381.
\end{acknowledgments}

\bibliography{Ziqi_1}

\end{document}